\documentclass[aps,preprintnumbers,amsmath,amssymb]{revtex4}
\usepackage{amscd}
\usepackage{amsmath}
\usepackage{amssymb}

\usepackage{float}
\usepackage{graphics,epsfig}
\usepackage{graphicx}
\usepackage{dcolumn}
\usepackage{bm}

\begin{document}
\baselineskip=0.8 cm
\title{{\bf Critical Dimension for Stable Self-gravitating Stars in $AdS$}}
\author{Zhong-Hua Li}
\email{sclzh888@163.com} \affiliation{Department of Physics, China
West Normal University, Nanchong 637002, China}

\author{Rong-Gen Cai}
\email{cairg@itp.ac.cn} \affiliation{Key Laboratory of Frontiers in
Theoretical Physics, Institute of Theoretical Physics, Chinese
Academy of Sciences, P.O.Box 2735, Beijing 100190, China}

\vspace*{0.2cm}
\begin{abstract}
\baselineskip=0.6 cm
\begin{center}
{\bf Abstract}
\end{center}

We study the self-gravitating stars with a linear equation of state,
$P=a \rho$, in AdS space, where $a$ is a constant parameter. There
exists a critical dimension, beyond which the stars are always
stable with any central energy density; below which there exists a
maximal mass configuration for a certain central energy density and
when the central energy density continues to increase, the
configuration becomes unstable.  We find that the critical dimension
depends on the parameter $a$, it runs from $d=11.1429$ to $10.1291$
as $a$ varies from $a=0$ to $1$. The lowest integer dimension for a
dynamically stable self-gravitating configuration should be $d=12$
for any $a \in [0,1]$ rather than $d=11$, the latter is the case of
self-gravitating radiation configurations in AdS space.
\end{abstract}
\maketitle
\newpage

\section{Introduction}

Self-gravitating configuration is a subject of long-standing
interest in general relativity since its thermodynamic behavior is
quite different from that of a usual thermal system without gravity,
due to the attractive long-range, unshielded nature of gravitational
potential. For example, it is well known that the canonical ensemble
is not defined in asymptotically flat space. This is because having
thermal radiation at constant temperature at infinity is not
compatible with asymptotic flatness. One can avoid this problem by
enclosing the system in a box, which is unphysical, or by working in
anti-de Sitter (AdS) space which needs not any unphysical perfectly
reflecting walls at finite radius, the rising gravitational
potential in AdS space, plus natural boundary conditions at
infinity, acts to confine whatever is inside. On the other hand, due
to the conjecture of AdS/CFT correspondence~\cite{Mald}, which says
that string theory/M theory on an AdS space (times a compact
manifold) is dual to a strong coupling conformal field theory (CFT)
residing on the boundary of the AdS space, over the past decade, a
lot of attention has been focused on AdS space and relevant physics.
Further, Witten~\cite{Witten} argued that thermodynamics of black
holes in AdS space can be identified to that of dual strong coupling
CFTs. Therefore one could study thermodynamics and phase structure
of strong coupling CFTs by investigating thermodynamics and phase
structure of AdS black holes. It is well-known that thermodynamics
of AdS black holes is quite different from that of their
counterparts in asymptotically flat space and that there is a
Hawking-Page phase transition between large black hole and thermal
radiation in AdS space~\cite{HP}. Therefore it is also of great
interest to study self-gravitating configurations in AdS space and
their thermodynamics.

Self-gravitating radiation gas has been investigated thoroughly.
Sorkin, Wald and Zhang~\cite{SWZ} have studied the equilibrium
configurations of self-gravitating radiation in a spherical box of
radius $R$ in asymptotically flat space. It was found that for
locally stable configuration, the total gravitational mass of
radiation obeys the inequality $M<\mu_{\rm max} R$, where $\mu_{\rm
max}=0.246$. In AdS space, Page and Philips~\cite{PP} examined the
self-gravitating configuration of radiation in four dimensional
space-time. The configuration can be labeled as its mass, entropy
and temperature versus the central energy density. They found that
there exist locally stable radiation configurations all the way up
to a maximum red-shifted temperature, above which there are no
solutions; there is also a maximum mass and maximum entropy
configuration occurring at a higher central density than the maximal
temperature configuration.  Beyond their peaks the temperature, mass
and entropy undergo an infinite series of damped oscillations, which
indicates the configurations in this regime are unstable. The
self-gravitating radiation in five dimensional AdS space has been
studied in \cite{HLR} (see also \cite{Hemm}) with similar
conclusions. Recently, Vaganov~\cite{Vaga} and
Hammersley~\cite{Hamm} independently discussed the self-gravitating
radiation configurations in higher dimensional AdS spaces. They
found that in the case of $4 \le d \le 10$, the situation is
qualitatively similar to the case in four dimensions, while $ d \geq
11$, the oscillation behavior disappears. Namely, there is a
critical dimension, $d_{c.ads}=11$ (very close, but not exact),
beyond which, the temperature, mass and entropy of the
self-gravitating configuration are monotonic functions of the
central energy density, asymptoting to their maxima as the central
density goes to infinity. The equilibrium configurations of
self-gravitating radiation gas in AdS space are quite different from
those of their counterparts in asymptotically flat space.

Recently Chavanis investigated spherically symmetric equilibrium
configurations of relativistic stars with a linear equation of state
$P=a\rho$ in flat space~\cite{Chav}. The equation of state $P=a\rho$
says the pressure $P$ is proportional to the energy density $\rho$,
where $a$ is a ratio coefficient. Theoretically, different
coefficients correspond to different stars, although it is
impossible to define names for all the stars corresponding to each
coefficient. The speed of sound is given by $(dP/d\rho)^2c=a^2c$,
where $c$ is the speed of light.  Thus causality requires
$a^2\leq1$. As a result, $0\leq a\leq1$ is considered only. One
could identify relativistic stars by an equation of state with their
different ratio coefficients. In the Newtonian limit $a\rightarrow
0$, it returns the classical isothermal equation of state. On the
other hand, the case with $a=1/(d-1)$, where $d$ is the dimension of
space-time, corresponds to a gas of self-gravitating radiation; the
core of neutron stars or so called ``photon stars" where the
pressure is entirely due to radiation. A gas of baryons interacting
through a vector meson field for the case of $a=1$ is called the
``stiffest" star. Chavanis found that the structure of the system is
highly dependent on the dimensionality of space-time and that the
oscillations in the mass-central density profile disappear above a
critical dimension $d_{c.flat}(a)$ depending on the coefficient $a$.
Above this dimension, the equilibrium configurations are stable for
$any$ central density, contrary to the case $d<d_{c.flat}$. For
Newtonian isothermal stars ($a\rightarrow 0$), the critical
dimension $d_{c.flat}(0)=11$. For the stiffest stars ($a=1$),
$d_{c.flat}(1)=10$ and for a self-gravitating
radiation($a=1/(d-1)$), $d_{c.flat}=1+9.96404372...$ very close to
$11$. The oscillations exist for any $a\in [0,1]$ when $d \le 10$
and they cease to exist for any $a\in [0,1]$ when $d\geq 11$.

In a previous paper~\cite{LHC}, we have studied self-gravitating
radiation configurations with plane symmetry in AdS space and found
that the situation is quite different from the one in spherically
symmetric AdS space. In the present paper we continue this study and
consider the self-gravitating configurations with a linear equation
of state $P=a\rho$ in higher ($d\geq4$)-dimensional spherically
symmetric AdS space, in order to see the dependence of the critical
dimension on the parameter $a$ and to see whether there is any
essential difference between the configurations in flat space and
the configurations in AdS space.

The organization of the paper is as follows. In the next section, we
give a general formulism to describe the relativistic stars with a
linear equation of state $P=a\rho$ in AdS space. The numerical
results are given in Sec.~III. The Sec.~IV is devoted to the
conclusions.

\section{relativistic stars with linear
equation of state in AdS space}

Consider a $d$-dimensional asymptotically AdS space with metric
\begin{equation}
\label{2eq1}
 ds^2= -e^{2\delta(r)} h(r) dt^2 + h^{-1}(r)dr^2 + r^2 \gamma_{ij}
 dx^idx^j,
 \end{equation}
 where $\delta$ and $h$ are two functions of the radial coordinate $r$,
 and $\gamma_{ij}$ is the metric of a $(d-2)$-dimensional Einstein manifold with constant
 scalar curvature $(d-2)(d-3)$. We take the gauge
 $\lim_{r\to \infty} \delta (r) =0$, and rewrite the metric function
 $h(r)$ as
 \begin{equation}
 \label{2eq2}
 h(r) = 1+ \frac{r^2}{l^2}- \frac{16\pi G m(r)}{(d-2) \Omega
 r^{d-3}},
 \end{equation}
 where $l$ denotes the radius of the AdS space with cosmological constant
 $\Lambda=-(d-1)(d-2)/2l^2$, $\Omega$ is the volume of the Einstein
 manifold,
  and $m(r)$ is the mass function of the
 solution. In this gauge, the total gravitational mass of the solution is
 just
 \begin{equation}
 M = \lim_{r\to \infty} m(r).
 \end{equation}
 The Einstein field equations with the cosmological constant and energy-momentum tensor
 $T_{\mu\nu}$ are
 \begin{equation}
 \label{2eq4}
 R_{\mu\nu}-\frac{1}{2} g_{\mu\nu}R -\frac{(d-1)(d-2)}{2l^2}g_{\mu\nu}= 8\pi G T_{\mu\nu}.
 \end{equation}
The metric function $\delta(r)$ and the mass
 function $m(r)$ satisfy the following equations
 \begin{eqnarray}
 \label{2eq9}
 && \delta'(r)=-\frac{8\pi G r}{(d-2) h(r)}\left(T^t_{\ t}-T^r_{\
 r}\right ), \nonumber \\
 && m'(r)=-\Omega r^{d-2} T^t_{\ t}.
 \end{eqnarray}
 The stress-energy tensor of perfect fluid is
 $T_{\mu\nu}=(\rho+P)U_{\mu}U_{\nu} +P g_{\mu\nu}$, where $U_{\mu}$
 is the four-velocity of the fluid. For self-gravitating radiation, the equation of state obeys $P =
 \rho/(d-1)$. For linear relativistic stars, $P=a\rho$ with $a \in [0,1]$.
 Thus the equations in (\ref{2eq9}) reduce to
 \begin{eqnarray}
 \label{2eq11}
&& \delta'(r)= \frac{8\pi G r}{(d-2)h(r)}(1+a) \rho ,  \\
\label{2eq12}
 && m'(r)=\Omega r^{d-2} \rho.
 \end{eqnarray}
In this paper we are particularly interested in the relation between
the total mass of configuration and the central energy density. To
integrate (\ref{2eq11}) and (\ref{2eq12}), we make a scaling
transformation as follows,
\begin{equation}
r \to l\,r, \ \ \ \rho \to l^{-2} \rho,  \ \ \  m(r) \to l^{d-3}
m(r),
\end{equation}
 so that $r$, $\rho$ and $m$ become dimensionless. In the numerical
 integration, we will adopt the units $8\pi G=1$ and
 $l=1$, and rescale the mass function as
 $$16\pi G m(r)/(d-2)\Omega \to m(r).$$
 In that case, the gravitational ``mass" $M$ in the plots in the
 next section in fact is the gravitational mass density, $16\pi G M/(d-2)\Omega$,
of corresponding relativistic star configurations. From the
conservation of the energy-momentum tensor,  one can derive the
energy density $\rho(r)$ satisfies the following equation
\begin{equation}
\label{2eq13} \frac{d\rho}{dr}=\frac{(a+1)\rho r}{2a}\left(
\frac{d-3}{r^2} -\frac{d-1}{h(r)}-
\frac{d-3}{r^2h(r)}-\frac{2a\rho}{(d-2)h(r)}\right).
\end{equation}
Further, note that the equations (\ref{2eq12}) and (\ref{2eq13})
 are singular at $r=0$. To avoid this, in the numerical calculations, we will
 start the integration from $r=\epsilon=10^{-5}$ to $r=L=100$. Obviously, the accuracy of the numerical
 calculations depends on the values of $\epsilon$ and $L$.

\section{Numerical results}

\begin{figure}
\includegraphics[width=10cm]{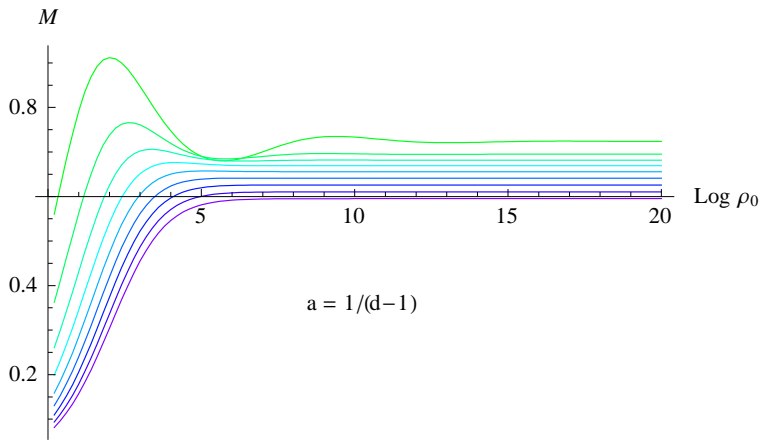}
\caption{\label{self1}The self-gravitating radiation in AdS space:
The mass of the self-gravitating radiation configuration versus the
central energy density from dimension $d=4$ (top curve) to $d=12$
(bottom curve).}
\end{figure}

\begin{figure}
\includegraphics[width=10cm]{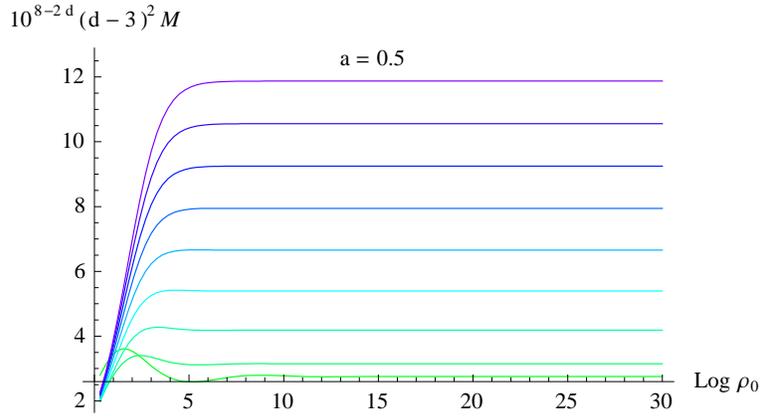}
\caption{\label{a=05n412}The case of $a=0.5$ in AdS space: The mass
of  relativistic star configurations versus the central energy
density from  dimension $d=4$ (bottom curve) to $d=12$ (top curve).}
\end{figure}

To be more clear to discuss all kinds of critical dimensions of
relativistic stars with a linear equation of state in AdS space, let
us first revisit the self-gravitating radiation case. In~\cite{Hamm}
 Hammersley found that in the case of $4\leq d\leq
10$, there exist locally stable radiation configurations all the way
up to a maximum mass, beyond their peaks the curves undergo an
infinite series of damped oscillations, which indicates the
configurations in this regime are unstable. However, with $d\geq
11$, the oscillation behavior disappears. The configurations turn to
be monotonic functions of the central energy density, asymptoting to
their maxima as the central density goes to infinity. Namely, there
is a critical dimension. By numerical analysis,
Hammersley~\cite{Hamm} found a semi-empirical model
\begin{equation}
\label{2eq14} \log\rho_{c}\approx
0.50d+\frac{5.75}{\sqrt{11.0-d}}-2.20,
\end{equation}
which gives a critical dimension $d_{c.ads}=11$, and a similar
conclusion was also reached by Vaganov~\cite{Vaga} independently. We
reproduce the result in Fig.~\ref{self1}.

\begin{figure}
\includegraphics[width=10cm]{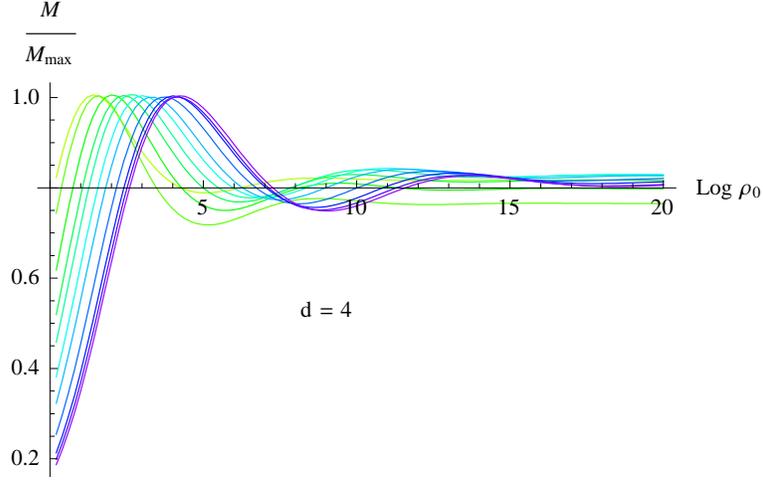}
\caption{\label{ad=4} The case of d=4 in AdS space: The mass of
relativistic star configurations versus the central energy density
from the coefficient  $a=1$ (left curve) to $a=0.01$ (right curve).}
\end{figure}

\begin{figure}
\includegraphics[width=10cm]{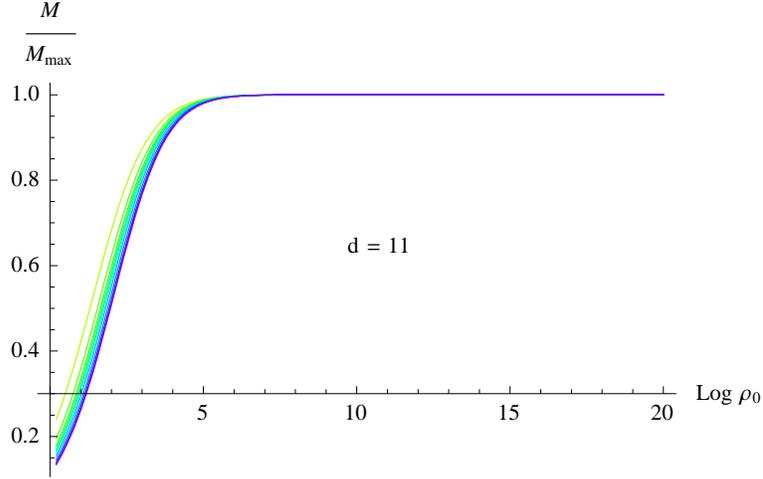}
\caption{\label{ad=11} The case of d=11 in AdS space: The mass of
 relativistic star configurations versus the central energy
density from the coefficient $a=1$ (left curve) to $a=0.01$ (right
curve).}
\end{figure}

Now we turn to the case of relativistic stars with a linear equation
of state $P=a\rho$.  Due to the requirement of causality, we
consider the cases with $0\leq a\leq1$. In Fig.~\ref{a=05n412}, we
plot the mass of relativistic stars configurations versus the
central energy density from dimension $d=4$ to $d=12$ for the case
of $a=0.5$. Note that in Fig.~\ref{a=05n412} we rescale the mass $M$
with scale $10^{8-2d}(d-3)^{2}$ in each dimension for better
displaying all curves in one figure. We notice that for a certain
value of $a$, the configurations of different dimensions from $4$ to
$12$ are similar to the case of self-gravitating radiation stars.
For the case of lower dimensions, there exist obvious oscillations
which can be seen from Fig~\ref{ad=4}. As for the higher dimensions,
the oscillations become weaker and weaker, and finally beyond some
critical dimensions, configurations become monotonic functions of
the central energy density. Fig.~\ref{ad=11} is the case of $d=11$.
Note that in Fig.~\ref{ad=4} and Fig.~\ref{ad=11} each curve
corresponds to each relativistic star with coefficient $a$ which
undergoes a continuous variation from 0.01 to 1.0 with step 0.1, and
also we rescale the mass density $M$ to $M/M_{\rm max}$ in each case
of coefficient $a$ for better comparing those different cases.

\begin{figure}
\includegraphics[width=10cm]{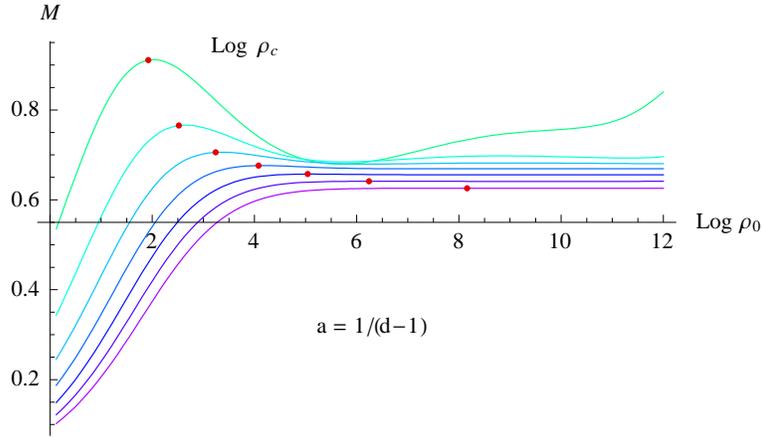}
\caption{\label{adskd1} The self-gravitating radiation in AdS space:
The mass of the self-gravitating radiation configuration versus the
central energy density from  dimension $d=4$ (top curve) to $d=10$
(bottom curve). The red dots are the saturation points
$\log\rho_{c}$.}
\end{figure}

As a matter of fact, it is not really accurate to determine the
critical dimensions only by the naked eyes to watch these curves. To
be more accurate to determine the critical dimensions, we take the
numerical analysis approach. In Fig.~\ref{adskd1} one can see
clearly that for each curve there exists a saturation point,
$\log\rho_{c}$ (the red dot in the figure), which is the location of
the first local maximum. The saturation point moves towards the
right side as the dimension $d$ increases. Beyond the saturation
points, the curves undergo an infinite series of damped
oscillations. When the dimension increases, the oscillations become
weaker and weaker, while the saturation point $\log\rho_{c}$ becomes
larger and larger, and finally, beyond a critical dimension
$d_{c.ads}$, the saturation point $\log\rho_{c}$ goes to infinity.
So we can determine the critical dimension by analyzing the
variation of the saturation points.

 Some data of saturation points $\log\rho_{c}$ are listed in Table
 ~\ref{tb1}. Obviously, $\log\rho_{c}$ depends on the coefficient $a$ and dimension
 $d$, so namely $\log\rho_{c}(a,d)$. We use a formula like (\ref{2eq14}) and obtain the
 critical dimensions $d_{c}(a)$ for different coefficient $a$,
 which are listed in Table~\ref{tb2}.
  As can be seen from Table~\ref{tb1},
 in the case of $d=11$ some data singularity appears.
So we only use data from $d=4$ to $d=10$ to fit.

\begin{table}[!h]
\tabcolsep 0pt \caption{\label{tb1}Saturation Points
$\log\rho_{c}(a,d)$} \vspace*{-12pt}
\begin{center}
\def\temptablewidth{0.5\textwidth}
{\rule{\temptablewidth}{1pt}}
\begin{tabular*}{\temptablewidth}{@{\extracolsep{\fill}}c|cccccccc}
a & d=4 &d=5 &d=6 &d=7 &d=8 &d=9 &d=10&d=11\\
\hline
      0.01&3.90&3.60&3.90&4.20&5.10&6.00&8.10&21.00\\
       0.1&3.00&3.00&3.30&3.90&4.80&6.00&8.40&21.90 \\
       0.2&2.40&2.40&3.00&3.90&4.80&6.30&8.70& $\infty$\\
       0.3&1.80&2.10&3.00&3.90&4.80&6.30&9.00& $\infty$\\
       0.4&1.50&2.10&3.00&3.90&5.10&6.60&9.60& $\infty$\\
       0.5&1.20&2.10&3.00&3.90&5.10&6.60&10.20&$\infty$\\
       0.6&1.20&2.10&3.00&4.20&5.10&6.90&11.10&$\infty$\\
       0.7&1.20&2.10&3.00&4.20&5.40&7.20&12.30&$\infty$\\
       0.8&1.20&2.10&3.30&4.20&5.40&7.50&14.10&$\infty$\\
       0.9&1.20&2.10&3.30&4.20&5.70&7.50&15.60&$\infty$\\
       1.0&1.20&2.10&3.30&4.20&5.70&7.80&16.50&$\infty$
       \end{tabular*}
       {\rule{\temptablewidth}{1pt}}
       \end{center}
       \end{table}

\begin{table}[!h]
\tabcolsep 0pt \caption{\label{tb2}Critical Dimensions
$d_{c.ads}(a)$} \vspace*{-12pt}
\begin{center}
\def\temptablewidth{0.5\textwidth}
{\rule{\temptablewidth}{1pt}}
\begin{tabular*}{\temptablewidth}{@{\extracolsep{\fill}}c|cccccc}
    a      & 0.01    &0.1     &0.2      &0.3    &0.4&0.5\\
\hline
$d_{c.ads}(a)$&11.1061 &11.0952 &10.9989  &10.9254&10.7465&10.6333\\
\hline
   a        &&0.6    &0.7     &0.8      &0.9&1.0\\
\hline
$d_{c.ads}(a)$&&10.4928 &10.3707 &10.2602  &10.2023&10.1763\\
           \end{tabular*}
       {\rule{\temptablewidth}{1pt}}
       \end{center}
       \end{table}

\begin{figure}
\includegraphics[width=10cm]{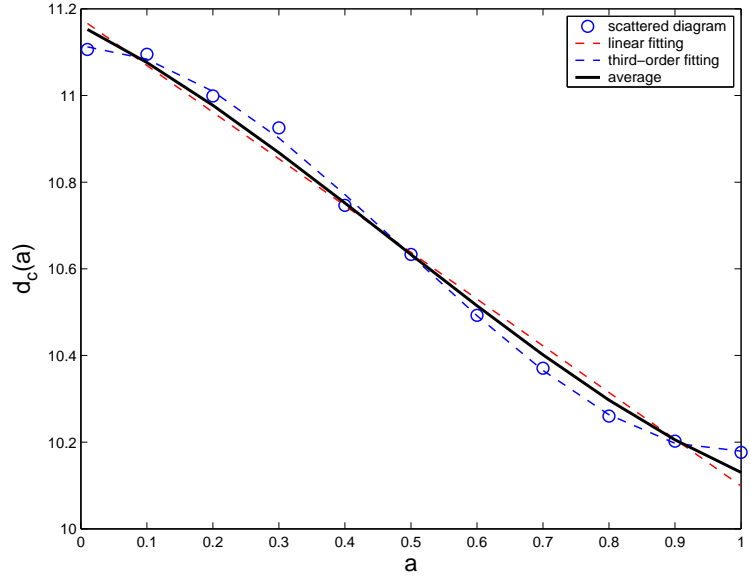}
\caption{\label{adcurve3}The critical dimensions $d_{c.ads}(a)$
versus the coefficient $a$. The blue hollow dots are the scattered
data of the critical dimensions $d_{c.ads}(a)$. The red dashed curve
is the case of linear fitting. The blue dashed curve is the case of
the third-order fitting. The black thick curve is the case of
average fitting.}
\end{figure}

\begin{figure}
\includegraphics[width=10cm]{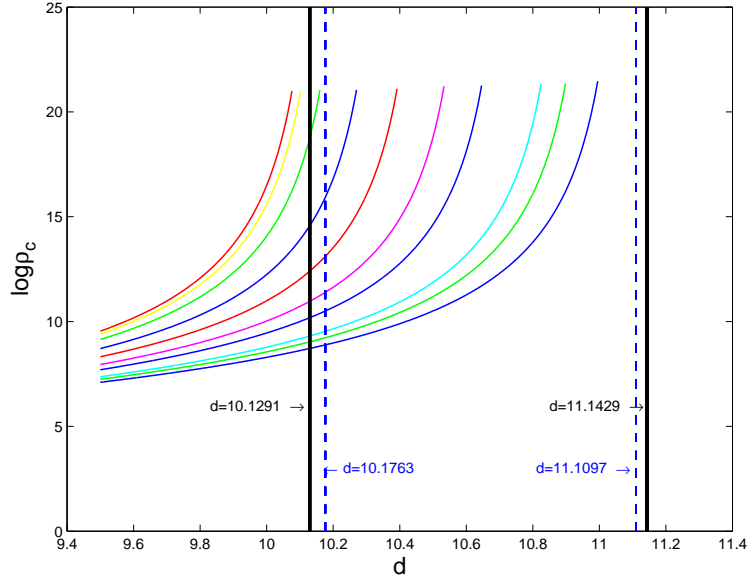}
\caption{\label{dcmaxmin}The saturation points $\rho_{c}$ versus the
dimension $d$ from $a=0.1$ (right blue curve) to $a=1.0$ (left red
curve). The blue dashed curves are the low and up bounds of the
critical dimensions $d_{c.ads}(a) \in [10.1763,11.1097]$,
respectively. The black thick curves are the low and up bounds of
the critical dimensions $d_{c.ads.aver}(a) \in [10.1291,11.1429]$,
respectively.}
\end{figure}

\begin{table}[!h]
\tabcolsep 0pt \caption{\label{tb3} Numerical Fitting Results}
\vspace*{-12pt}
\begin{center}
\def\temptablewidth{0.5\textwidth}
{\rule{\temptablewidth}{1pt}}
\begin{tabular*}{\temptablewidth}{@{\extracolsep{\fill}}c|cccc}
   $d_{c.ads}(a)$  & linear & second-order  &third-order& average\\
\hline
low bound &10.1063&10.1048&10.1763& 10.1291\\
\hline
 up bound &11.1601&11.1591  &11.1097& 11.1429\\
           \end{tabular*}
       {\rule{\temptablewidth}{1pt}}
       \end{center}
       \end{table}
In Fig.~\ref{adcurve3} we plot the scattered diagram of the critical
dimensions $d_{c.ads}(a)$ versus the coefficient $a$. It can be seen
that the value of $d_{c.ads}(a)$ becomes smaller and smaller as the
value of the coefficient $a$ goes from $0$ to $1$. So $d_{c.ads}(a)$
can be considered to be a monotonic function of the coefficient $a$.
The red dashed line in Fig.~\ref{adcurve3} is the result of a linear
numerical fitting for these scattered data. It could be seen that
for all kinds of relativistic stars with a linear equation of state
$P=a\rho$ $(a\in[0,1])$, the critical dimensions $d_{c.ads}(a) \in
[10.1063,11.1601]$. Obviously, the lower and upper bounds of
$d_{c.ads}(a)$ for this kind of fitting are a little wider because
these scattered data are not really linear related. To obtain more
accurate results, we need to adopt non-linear fitting which could
better fit the scattered data. The results of the second and
third-order fitting are listed in Table~\ref{tb3}. The blue dashed
curve in Fig.~\ref{adcurve3} is the case of the third-order fitting.
Note that the curve of the second-order fitting is not plotted in
Fig.~\ref{adcurve3} because it is overlapped with the most of the
linear fitting curve. It can be seen from Fig.~\ref{adcurve3} that
the curve of the third-order fitting can better fit the scattered
data. Namely, in AdS space for all kinds of relativistic stars with
a linear equation of state $P=a\rho$ $(a\in[0,1])$ the critical
dimensions $d_{c.ads}(a) \in [10.1763,11.1097]$. The minimum for the
stiffest star in AdS space is $d_{c.ads}(1)=10.1763$ and the maximum
for the Newtonian isothermal star is $d_{c.ads}(0)=11.1097$. In
Fig.~\ref{dcmaxmin} we plot the saturation points $\log\rho_{c}$
versus the dimension $d$ for each relativistic star (corresponding
to each coefficient $a$ ) from $a=0.1$ (right blue curve) to $a=1.0$
(left red curve) in the range $ 9.4\le d \le 11.4$. The low and up
bounds of the critical dimensions $d_{c.ads}(a) \in
[10.1763,11.1097]$ are also plotted in this figure (the blue dashed
curves). One can find that for the stiffest stars $(a=1)$, the left
red curve increases steeply and finally up to a critical dimension
$d=10.1763$. However, for Newtonian isothermal stars
$a\rightarrow0$, the right blue curve changes more slowly, up to the
critical dimension $d=11.1097$.

By an analysis of asymptotic behavior, Chavanis~\cite{Chav} found
that the critical dimension runs from $d=10$ to $11$ when the
coefficient $a$ decreases from $a=1$ to $0$ in flat space. Namely
the difference of dimension $\triangle
d_{flat}=d_{c.flat}(0)-d_{c.flat}(1)=1$. In our case, namely in AdS
space, the critical dimension runs from $d=11.1097$ to $10.1763$
with $\triangle d_{ads}=d_{c.ads}(0)-d_{c.ads}(1)=0.9334$ when $a$
varies from $0$ to $1$. If taking the averaged values, then one has
$\overline{\triangle
d}_{ads}=d_{c.ads.aver}(0)-d_{c.ads.aver}(1)=11.1429-10.1291=1.0138$.
There exists some difference between the flat and AdS cases.
Therefore we guess that for all kinds of  relativistic stars with a
linear equation of state  regardless of in AdS and flat spaces, the
critical dimensions $d_{c.ads}(a)$ and $d_{c.flat}(a)$ vary just
within $1$ dimension when $a$ changes from $0$ to $1$, and the
inaccuracy between the two cases is very small, within $1.38\%$.
Therefor the critical dimension for a stable relativistic stars with
a linear equation of state is $12$, rather than $11$, which is the
case of self-gravitating radiation configurations.

\section{Conclusion}

There exists a critical dimension for self-gravitating
configurations in general relativity, beyond which the
configurations with any central energy density are always stable;
below which there exists a maximal mass configuration for a certain
central energy density, when the central energy density increases,
the configuration becomes unstable.  In this paper we studied the
self-gravitating configurations (relativistic stars) with a linear
equation of state $P=a \rho$ in AdS space, where $a$ is a constant
parameter within $a \in [0,1]$. We found that the critical dimension
depends on the parameter $a$, it runs from $d=11.1097$ to $10.1763$
in the third order fitting as $a$ varies from $a=0$ to $1$. The
result is a little different from the case in flat space. It runs
from $d=11.1429$ to $10.1291$ if one takes the averaged fitting. In
that case, it is very close to the case in flat space within the
inaccuracy $1.38\%$. Therefore it is of interest to study the
self-gravitating configurations in de Sitter space and to see
whether there exist any differences among the three cases.

In \cite{Vaga} and \cite{Hamm}, it was found that the critical
dimension of self-gravitating radiation configurations is $d=11$.
Combing the results obtained in \cite{LHC} and in the present paper,
we see that in fact, the dimension $d=11$ might not have any other
special physical meaning, except for the stability divide. Note that
the fact that no special happens for thermodynamics of AdS
Schwarzschild black holes in $d \ge 4$, it is of interest to
understand the meaning of the existence of the critical dimension in
the AdS/CFT correspondence.

\begin{acknowledgments}
This work was supported partially by grants from NSFC, China (No.
10821504 and No. 10525060), a grant from the Chinese Academy of
Sciences with No. KJCX3-SYW-N2.
\end{acknowledgments}

\end{document}